\documentclass[reprint,amsmath,amssymb,aps,pra]{revtex4-2}

\usepackage{upgreek} 
\usepackage{epsfig, amsmath,amsfonts, amssymb,graphicx,amsthm,color}
\usepackage{mathtools}
\usepackage{dcolumn}
\usepackage{bm}
\usepackage{rotating}
\usepackage{units} 
\usepackage{amssymb}
\usepackage{sidecap}
\usepackage{multirow}
\usepackage[american]{babel}
\usepackage{url}
\usepackage[colorlinks=true,        
            linkcolor=blue,         
            citecolor=blue,        
            urlcolor=blue,          
            pdfborder={0 0 0},      
]
            {hyperref}

\begin{document}

\title{Non-Dichroic Enantio-Sensitive Chiroptical Spectroscopy}
\author{Letizia Fede$^{1}$, Debobrata Rajak$^{2}$, Chris Sparling$^{3}$, David Ayuso$^{4}$, Val\'erie Blanchet$^{1}$, Piero Decleva$^{5}$, Dominique Descamps$^{1}$, St\'ephane Petit$^{1}$, Bernard Pons$^{1}$}
\author{ Yann Mairesse$^{1}$}
\email[]{yann.mairesse@u-bordeaux.fr}
\author{Andr\'es Ord\'o\~nez$^{6}$}
\email[]{andres.ordonez@fu-berlin.de}
\affiliation{$^{1}$ Universit\'e de Bordeaux -- CNRS -- CEA, CELIA, UMR5107, Talence, France}
\affiliation{$^{2}$ ELI ALPS, The Extreme Light Infrastructure ERIC, Wolfgang Sandner u. 3., 6728 Szeged, Hungary }
\affiliation{$^{3}$ Institute of Photonics and Quantum Sciences, Heriot-Watt University, Edinburgh EH14 4AS, UK}
\affiliation{$^{4}$ Department of Chemistry, Molecular Sciences Research Hub, Imperial College London, SW7 2AZ London, UK}
\affiliation{$^{5}$ Università degli Studi di Trieste, Trieste, Italy}
\affiliation{$^{6}$ Department of Physics, Freie Universit\"at Berlin, 14195 Berlin, Germany}

\begin{abstract}
Chiroptical effects using circularly polarized light produce signals that change sign when switching either molecular handedness (enantiosensitivity) or the light helicity (circular dichroism). Here, we break this enantiosensitive-and-dichroic paradigm by measuring a new type of chiroptical signal which is enantiosensitive but not dichroic. We photoionize chiral molecules using a strong laser field and detect the three-dimensional photoelectron momentum distribution. The non-dichroic, enantiosensitive asymmetry is encoded in octupolar and higher multipolar terms in the photoelectron angular distribution, which appear in multiphoton ionization with elliptically polarized fields or cross polarized two-color fields. The robustness of the enantiosensitivity with respect to the relative phase between the vectorial components of the ionizing field represents an example of symmetry protection, and opens unexplored opportunities for imaging ultrafast dynamics in chiral molecules, such as enantiosensitive photoelectron spectroscopy with bright squeezed vacuum states.  
\end{abstract}

\maketitle

Chiral molecules are well-known both for their importance in chemistry and biology as well as for their nuanced interaction with light's polarization, which leads to a broad range of chiroptical effects, in particular circular dichroisms \cite{berovaComprehensiveChiropticalSpectroscopy2012a}.  These effects find applications in the analysis and control of chiral molecules, as well as in the control of light itself. It is common knowledge in chiroptical spectroscopy that switching the handedness of the studied molecules or the ellipticity of the light illuminating them is equivalent. This equivalence was recently predicted to be broken in a specific configuration, namely the two-photon ionization of chiral molecules by a linearly polarized ultraviolet laser field and its orthogonally polarized second harmonic \cite{ordonez2022}. The three-dimensional (3D) momentum distribution of the ejected photoelectrons was predicted to contain a component that switches sign when exchanging the molecular enantiomer, but not when switching the ellipticity of the ionizing light. The emergence of such Non-Dichroic EnantioSensitive (NoDES) signals thus represents a new paradigm in chiroptical spectroscopy. The existence of NoDES was up to now predicted only in a specific optical configuration, and the magnitude of the effect was not evaluated, raising the question of the practical applicability of this new type of spectroscopy. 

In this Letter, we generalize the conditions required to observe NoDES and report experiments using both elliptically polarized and orthogonally polarized two-color laser fields. Our results demonstrate that NoDES signals in the 1$\%$ range can be resolved in 3D electron momentum distributions, but also in judiciously selected 2D projections of these 3D distributions. This letter is complemented by a companion article \cite{companionPRA} describing the theoretical framework of NoDES, and providing numerical calculations of NoDES signals in various configurations.

\begin{table}

\begin{centering}
  \begin{tabular}{l l}
    \hline
    Property & $(l,m)$ condition\tabularnewline
    \hline
    Symmetry allowed & even $m$\tabularnewline
    Enantiosensitive & odd $l$\tabularnewline
    Dichroic & (odd $l$ and $m\geq0$) or \tabularnewline
    & (even $l$ and $m<0$)\tabularnewline
    \hline
  \end{tabular}
 \end{centering}
\caption{Selection rules for $b_{l,m}$ coefficients [Eq. (\ref{eq:PAD})] resulting from ionization of randomly oriented chiral molecules 
with elliptical fields $\vec{E}(t)=E[\cos(\omega t)\hat{x}+\varepsilon\sin(\omega t)\hat{y}]$.}\label{tab:elliptical-selection-rules}
\end{table}

\begin{figure*}
\begin{centering}
\includegraphics[width=\linewidth]{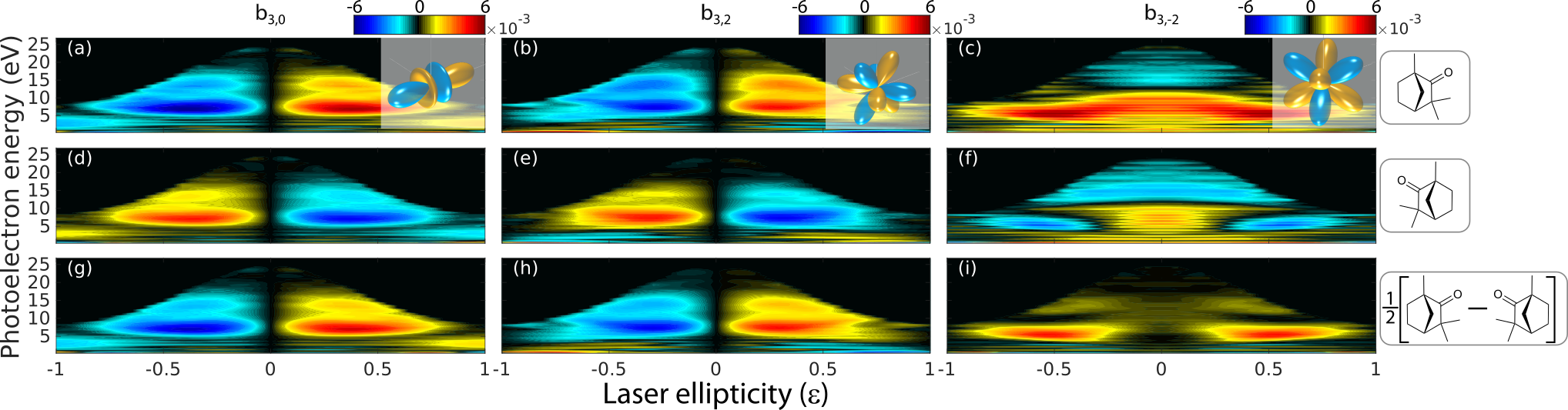}
\end{centering}
\caption{Spherical harmonic decomposition of the forward/backward asymmetric part of the 3D photoelectron angular distribution obtained by photoionizing (+)-fenchone (a,b,c) and ($-$)-fenchone (d,e,f) with an elliptically polarized 1030 nm laser field at $2\times10^{13}$ $\mathrm{W/cm^{2}}$. The plotted coefficients are $b_{3,0}$ (a,d,g), $b_{3,2}$ (b,e,h) and $b_{3,-2}$ (c,f,i), normalized by the yield $b_{0,0}$ at each photoelectron energy. (g-i) show the results obtained by subtracting the response of opposite enantiomers.}
\label{figSelecta}
\end{figure*}

The photoionization of an isotropic target by absorption of $N$ photons produces a photoelectron angular distribution (PAD) that can be described as a sum of real spherical harmonics:
\begin{equation}
  P(\theta,\phi)=\sum_{l=0}^{2N}\sum_{m=-l}^l b_{l,m}Y_{l,m}(\theta,\phi).\label{eq:PAD}
\end{equation}
  The contribution of the different spherical harmonics is determined by selection rules that have their origin in symmetries, which can be of geometric or dynamic character. Let us consider photoionization by an elliptically polarized laser field. A symmetry analysis analogous to that in Ref. \cite{ordonez2022} and detailed in \cite{companionPRA}, considering how the elliptical field, the isotropic sample, and the real spherical harmonics change upon rotations and reflections, leads to table \ref{tab:elliptical-selection-rules}, which shows which $b_{l,m}$ coefficients are symmetry allowed and whether they are dichroic (i.e. change sign with ellipticity) and/or enantiosensitive (change sign with enantiomer).

In circular polarization, where $b_{l,m\neq0 }=0$, the odd-$l$ terms reflect the forward/backward asymmetry (FBA) of the PAD, referred to as PhotoElectron Circular Dichroism (PECD) \cite{ritchie1976,powis2000,sparling2025}. When the ionizing radiation is elliptically polarized, nonzero $b_{l,m\neq0 }$ terms can emerge. For instance, a significant $b_{3,2}$ contribution was recently measured in the photoelectron elliptical dichroism (PEELD) signal resulting from the three-photon ionization of camphor \cite{sparling2023a}. This term was found to be enantiosensitive and dichroic, which is consistent with Table \ref{tab:elliptical-selection-rules}.

Here, we aim at revealing the enantiosensitive non-dichroic coefficients which, according to Table \ref{tab:elliptical-selection-rules}, have odd $l$ and $m<0$. No significant contribution of such terms was reported in \cite{sparling2023a}. In order to favor their emergence, we perform our investigation in the above-threshold ionization regime \cite{agostini1979}, in which elliptically dichroic azimuthal offsets of the PAD, characteristic of $m<0$ terms, were observed in atoms \cite{bashkansky1988,muller1988,lambropoulos1988,nikolopoulosAbovethresholdIonizationNegative1997,manakovEllipticDichroismAngular1999,wangDeterminationCrossSections2000,borcaThresholdEffectsAngular2001} and 
molecules \cite{davino2025}. In the strong-field ionization regime, these azimuthal offsets result from the influence of the ionic potential in the angular streaking of the ejected attosecond electron wavepacket by the rotating laser field \cite{becker2022,basile1988,goreslavski2004,popruzhenko2008,torlina2015,sainadh2019}, as well as from possible contributions of tunneling time delays \cite{eckle2008,landsman2014,camus2017,kheifets2020}.

We photoionized enantiopure fenchone molecules with 1030 nm, 130 fs pulses at 1 MHz repetition rate (Tangerine Short Pulse, Amplitude), focused at $2.5\times10^{13}$ $\mathrm{W/cm^{2}}$ in a Velocity Map Imaging (VMI) spectrometer collecting a 2D projection of the 3D PAD \cite{eppink1997}. The laser pulses propagated along the $z$ axis. The laser ellipticity $\varepsilon$ was continuously varied by rotating a zero-order half-wave plate in front of a fixed zero-order quarter-wave plate, while filming the phosphor screen of the VMI with a 16-bit camera. Each image was accumulated over $5\times10^{4}$ laser shots. The resulting film was Fourier-analyzed to extract the noise-free projection of the 3D PAD as a function of ellipticity, following the procedure described in \cite{rajak2024}. This process was repeated after rotating the quarter-wave plate by steps of $5^{\circ}$, to record 72 projections of the 3D PAD at each ellipticity, in each enantiomer. The 3D PAD was then tomographically reconstructed \cite{wollenhaupt2009}. In order to minimize the influence of inhomogeneities of the detector, PECD and PEELD signals are generally obtained by taking the difference of the signals obtained at opposite ellipticities. Here, since we measure a 3D distribution, we perform a mirror operation with respect to the $(x,z)$ plane before taking the difference of the signals, as detailed in \cite{bloch2021,rajak2024}.
\begin{figure*}
\begin{centering}
\includegraphics[width=0.75\linewidth]{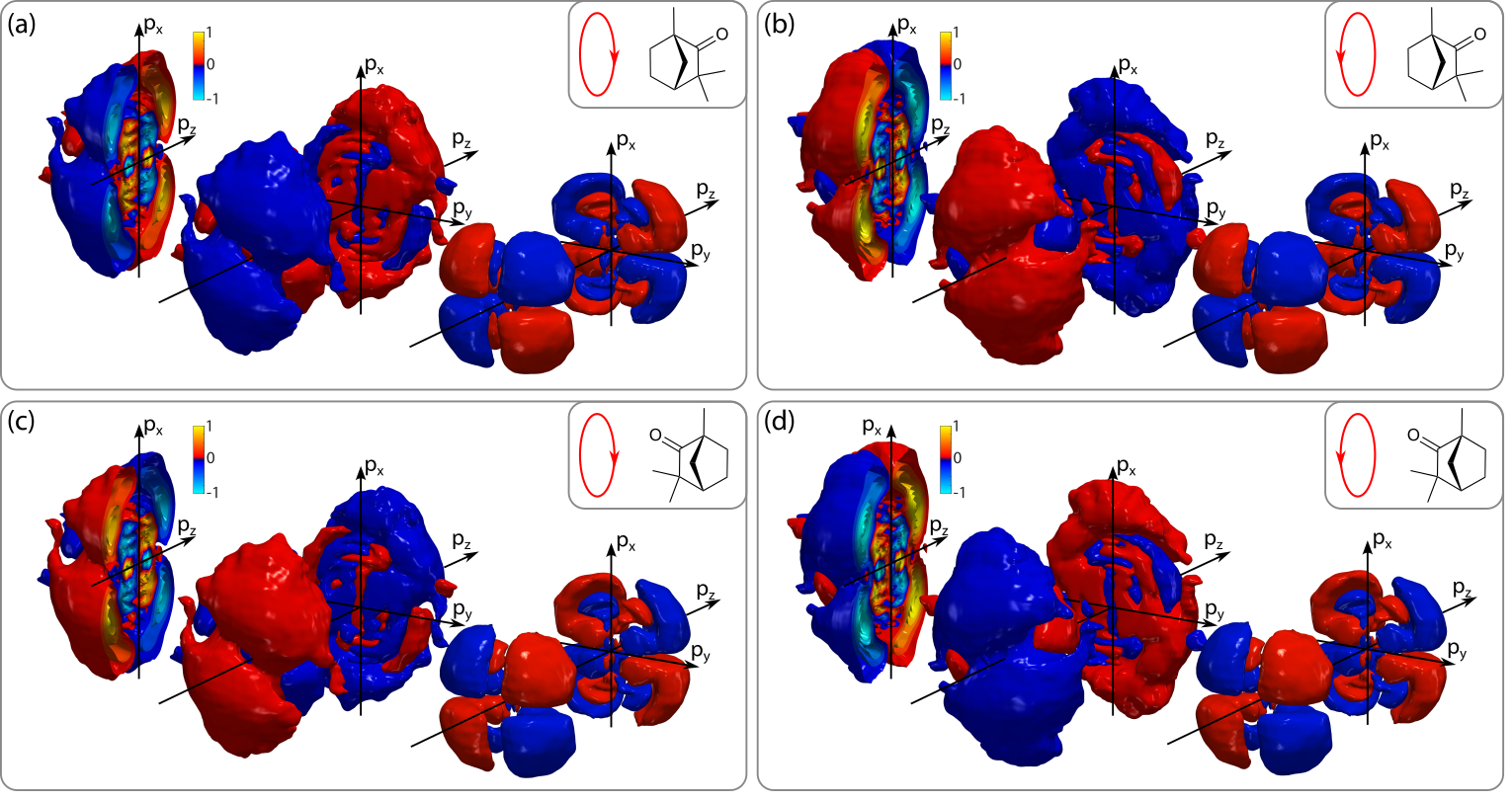}
\end{centering}
\caption{Forward-backward asymmetry (central plot in each panel) in the photoelectron angular distribution for (+)- (a,b) and ($-$)-fenchone (c,d), and clockwise (a,c) and counterclockwise ellipticities (b,d). The left plot in each panel shows a cut of the 3D FBA in the $(p_x, p_z)$ plane, while the right part is the NoDES distribution, obtained by antisymmetrizing the FBA along the $p_y$ axis.}
\label{fig3D}
\end{figure*}

We decomposed the resulting distribution in spherical harmonics using a least square fitting procedure \cite{politis2016}, up to maximum order $l=9$, and confirmed the results by using the FHANTOM algorithm \cite{sparling2024}. To illustrate the different behavior of the $b_{l,m}$ coefficients, Fig. \ref{figSelecta} shows the evolution of $b_{3,0}$ (a,d,g), $b_{3,2}$ (b,e,h) and $b_{3,-2}$ (c,f,i), normalized by the yield $b_{0,0}$, as a function of the kinetic energy of the photoelectron and the laser ellipticity. The other coefficients are shown in the End Matter section. The shape of the corresponding spherical harmonics are displayed in the insets. The results are for (+)-fenchone (a-c) and ($-$)-fenchone (d-f). The enantiodifferential response, obtained by as half the difference of the results from (+)- and ($-$)-fenchone, are depicted in Fig. \ref{figSelecta}(g-i).

The $b_{3,0}$ coefficient shows the characteristic evolution of a PEELD signal \cite{comby2018a}. In circular polarization, it maximizes around 3 eV photoelectron energy, and changes sign when switching ellipticity or enantiomer. As the magnitude of the laser ellipticity $|\varepsilon|$ decreases, the photoelectron spectrum extends to higher electron energies, indicating that electron rescattering takes place, leading to photoelectron energies up to 10 $U_p$ in linear polarization, where $U_p\sim2.5$ eV is the ponderomotive potential \cite{paulus1994}. The $b_{3,0}$ coefficient maximizes around  $|\varepsilon| = 0.4$, reaching $0.6\%$ at a photoelectron kinetic energy of 7 eV, and shows a secondary maximum ($0.25\%$) at 13 eV.  The $b_{3,2}$ coefficient shows a similar behavior, but vanishes in circular polarization, as expected for a cylindrical symmetry. $|b_{3,2}|$ maximizes at slightly lower ellipticity, around $|\varepsilon|= 0.3$, and shows a sign change in the low energy range, at an energy that depends on the ellipticity. This reflects the interference of direct and rescattered electrons in this energy range. These results demonstrate that $b_{3,0}$ and $b_{3,2}$ are enantiosensitive and dichroic, as observed in 3-photon ionization of camphor \cite{sparling2023a}.

Figures \ref{figSelecta} (c,f,i) show that the evolution of  $b_{3,-2}$ with $\varepsilon$ is very different. In (+)-fenchone, $b_{3,-2}$  reaches a maximum of $\sim 0.6\%$ at 5 eV for $|\varepsilon|=0.5$, while it peaks at $\sim-0.4\%$ at the same energy and ellipticity in ($-$)-fenchone. This shows that $b_{3,-2}$ changes sign when switching enantiomer, but not when switching the ellipticity of the ionizing field: it is enantiosensitive but non-dichroic. Figures \ref{figSelecta}(c,f) show an imperfect mirroring between enantiomers, as well as non-zero values of $b_{3,-2}$ for linearly polarized light. This reflects artifacts in the photoelectron imaging, resulting from the inhomogeneity of the detector as well as from possible stray magnetic fields in the VMI spectrometer. Taking the enantio-differential response (Fig. \ref{figSelecta}(i)) minimizes the effect of these artifacts. The resulting  $b_{3,-2}$ maximizes at $\sim 0.5\%$ at 5 eV for $|\varepsilon|=0.55$ and shows a secondary maximum of $\sim 0.1\%$ at 13 eV for $|\varepsilon|=0.37$.

\begin{figure*}
	\centering
	\includegraphics[width=\linewidth]{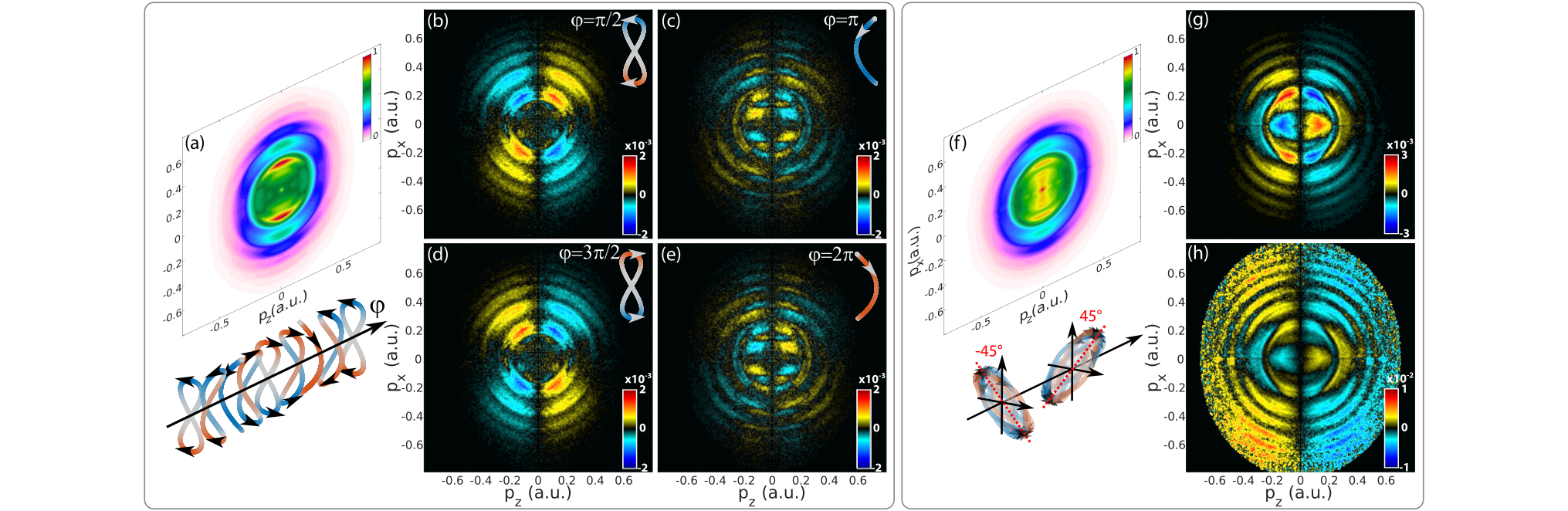} 
	\caption{Strong-field ionization of (+)-fenchone by an orthogonal two-color laser field. (a) Projected PAD, symmetrized along $p_{x}$ and $p_z$, and evolution of the FBA with $\varphi$ (b-e) when the fundamental field is polarized along $x$. (f) Projected PAD when the laser field polarization is rotated by $45^\circ$. (g) NoDES signal obtained by subtracting the FBAs recorded with $\pm 45^\circ$ polarization rotations of the field. (h) Normalized NoDES signal.}
	\label{FigXPOTC}
\end{figure*}

These results demonstrate the existence of NoDES terms in the 3D PAD produced by photoionizing chiral molecules by a strong elliptical laser field. Remarkably, these terms are similar in magnitude as, or even higher than, the other $m$ components of the spherical harmonic decomposition. The results, analyzed up to $l=9$ in the End Matter section, are in agreement with the symmetry-based selection rules of Table \ref{tab:elliptical-selection-rules}. They show that NoDES terms are present at all odd-$l$ orders, and highlight the interest of using strong-field ionization to obtain highly structured and sensitive 3D photoelectron distributions. To assess the structural sensitivity of NoDES signals, we repeated the measurements in $\alpha$-pinene molecules. The results, shown in End Matter, confirm the selection rules, and show a very high sensitivity of NoDES terms to the molecular structure. 

As all the spherical harmonics contributing to the NoDES signal are antisymmetric along $p_y$, we obtain the 3D NoDES  distribution directly from the 3D FBAs. Figure \ref{fig3D} shows the FBA of (+)- and ($-$)-fenchone at opposite ellipticities $\varepsilon=\pm 0.5$. The cuts in the ($p_x,p_z$) plane reveal the typical behavior of PEELD in strong fields, with characteristic high-momentum branches corresponding to the laser-induced diffracted electrons \cite{ray2008,blaga2012,rajak2024}. These electrons are preferentially diffracted forward in (+)-fenchone and backward in ($-$)-fenchone, when $\varepsilon=0.5$. Note that the responses of the two enantiomers have been averaged to produce artifact-free 3D distributions, such that the FBAs from (+)-fenchone (Fig. \ref{fig3D}(a,b)) and ($-$)-fenchone (Fig. \ref{fig3D}(c,d)) are mirror images of each other with respect to the polarization plane $(p_x, p_y)$ for a given ellipticity. 

The 3D distribution shows that the FBA is tilted (counter-) clockwise for (counter-) clockwise ellipticity, reflecting the angular streaking of the ionized electrons by the rotating laser field. For (+)-fenchone and clockwise polarization, this tilt induces an excess of electrons in the $(p_x>0,p_y>0,p_z>0)$ octant, see Fig. \ref{fig3D}(a) middle. This excess appears as a positive lobe in the NoDES distribution, obtained by extracting the antisymmetric component of the FBA along $p_y$, see Fig. \ref{fig3D}(a) right. For the opposite ellipticity, the combined sign change and opposite angular streaking of the FBA induces an excess of electrons in the same $(p_x>0,p_y>0,p_z>0)$ octant, see Fig. \ref{fig3D}(b) middle. Thus, the NoDES distribution remains unchanged, see Fig. \ref{fig3D}(b) right. By contrast, switching enantiomer reverses the sign of the NoDES, see Fig. \ref{fig3D}(c,d) right. Therefore, these pictures enable us to capture the origin of the NoDES terms in the 3D momentum distribution of photoelectrons as the combined action of the angular streaking by the laser field and the forward/backward asymmetric scattering in the chiral potential \cite{fehre2019b,bloch2021}.

The octupolar or higher order multipolar symmetry of the NoDES signals make them vanish in projected photoelectron angular distributions along $p_y$, which is the most commonly detected signal in VMI measurements. This is what led us to perform complete tomographic imaging of the 3D PAD. However, NoDES signals should be directly measurable in projections along specific directions such as  $\mathbf{\hat{x}+\hat{y}}$. Indeed, this configuration was proposed in \cite{ordonez2022}, for the two-photon ionization by a linearly polarized field and its orthogonally polarized second harmonic $\vec{E}(t)=E_{\omega}\cos(\omega t)\hat{x} + E_{2\omega}\cos(2\omega t + \varphi)\hat{y}$. The analysis of selection rules and numerical simulations for this configuration show that NoDES should be observable in both multiphoton and strong-field ionization \cite{companionPRA}.

We thus performed an experiment aiming at resolving the NoDES contribution in the strong-field ionization of fenchone molecules by a 1030-515 nm orthogonal two-color laser field. The fundamental laser beam at $\omega$ (50 W, 140 fs, 166 kHz) was split in two by a polarizing beamsplitter at the entrance of a Mach-Zehnder interferometer. In one arm, a 1 mm thick BBO crystal was used to frequency double the beam, and two dichroic mirrors were used to remove the fundamental component. The $\omega$ and $2\omega$ beams, with parallel polarizations, were recombined by a beamsplitter. A true zero-order dual-wavelength wave plate was adjusted to rotate the polarization direction of the fundamental beam by 90$^{\circ}$ while keeping the second harmonic polarization fixed (along $y$). The polarization direction of the resulting orthogonal two-color field was controlled by a super-achromatic half-wave plate in a motorized rotation stage, followed by a synchronized rotating 1.6 mm calcite plate ensuring perfectly linear and orthogonal polarizations of the fields by temporally delaying the polarization imperfections. Finally, the relative phase between the two colors $\varphi$ was controlled by translating a pair of SiO$_{2}$ wedges. The beams were focused at around $10^{13}$ W/cm$^2$, with an intensity ratio of the second harmonic to fundamental of $5\%$.

The measurements were performed by continuously acquiring 2D projections of the 3D PAD while scanning $\varphi$ over 18$\pi$ radians. The resulting film was Fourier-transformed to isolate the continuous component, i.e. the non-dichroic part of the signal that is insensitive to $\varphi$, as well as the component periodically oscillating with $\varphi$. The acquisition was repeated in opposite enantiomers to extract the average enantiodifferential signal. Figure \ref{FigXPOTC} shows the results of measurements where the polarization direction of the fundamental field is parallel to the VMI detector plane. The projected PAD (Fig. \ref{FigXPOTC}(a)) shows 6 rings corresponding to 6 above-threshold-ionization peaks, maximizing along the $p_z=0$ direction, i.e. in the laser polarization plane. The combination of fundamental and orthogonal second harmonic produces a field whose rotation direction reverses every half cycle of the fundamental, and whose shape depends on the relative phase $\varphi$, as depicted in Fig. \ref{FigXPOTC}(a). Despite their zero net ellipticity, these fields were shown to produce enantiosensitive forward/backward asymmetries in chiral photoionization \cite{demekhin2018,rozen2019}, with opposite signs in the upper and lower hemispheres, and which reverse when shifting $\varphi$ by $\pi$. This enantiosensitive and dichroic signal can be isolated by extracting the oscillating part of the FBA, shown in Fig. \ref{FigXPOTC}(b-e) for different $\varphi$. The distributions are almost perfectly up/down antisymmetric, reflecting the symmetry of the ionizing field. The FBA lies in the $0.2\%$ range of the peak PAD, and changes shape as a function of $\varphi$, because of the change in the shape of the ionizing field, as reported in \cite{rozen2019}.

Symmetry analysis as well as theoretical calculations show that NoDES components are produced by the orthogonal two-color laser field \cite{companionPRA}, but their antisymmetry along the second harmonic field polarization direction make them invisible in the recorded $p_y$-integrated 2D projections. To reveal them, we need to rotate the field polarization away from the integration axis of the VMI. We thus repeated the experiment for two other orientations of the ionizing laser field, rotated by $45^\circ$ and $-45^\circ$ with respect to the $y$ direction (Fig. \ref{FigXPOTC}(f)). We isolate the continuous ($\varphi$-independent) component of the Fourier transform of the signal, and subtract the FBA measured for $45^\circ$ and $-45^\circ$ projections, and for opposite enantiomers. The resulting signal, divided by 4 to obtain the average FBA in a single projection for a single enantiomer, is depicted in Fig. \ref{FigXPOTC}(g). It shows a clear non-dichroic, enantiosensitive signal. The NoDES is remarkably symmetric along the $p_{x}$ direction, confirming the predictions of \cite{ordonez2022} as well as the theoretical calculations of \cite{companionPRA}. The central ATI peak shows a sign switch of the NoDES signal as a function of the electron momentum and ejection angle, which qualitatively agrees with the shape of the projection of an octupolar distribution \cite{ordonez2022}. Higher ATI peaks show a more uniform NoDES signal. Normalizing the NoDES signal by the projected PAD  provides a signal in the $1\%$ range (Fig. \ref{FigXPOTC}(h)), which is the typical magnitude of PECD signals in the strong-field regime. The NoDES signal produced by the combination of orthogonally linearly polarized fields is thus remarkably strong. 

The observation of NoDES signals represents a new form of chiroptical spectroscopy, in which the ellipticity of the ionizing radiation does not have to be defined to discriminate enantiomers. This behavior is related to the concept of symmetry protection, which has been recognized in diverse physical systems, such as bound states in the continuum in photonic lattices \cite{plotnikExperimentalObservationOptical2011} or  topological phases in condensed matter \cite{guTensorentanglementfilteringRenormalizationApproach2009}. It reflects the principle that an effect remains robust as long as the relevant symmetries are maintained. NoDES signals provide a clear example of symmetry protection in the context of chiroptical spectroscopy -- they correspond to enantiosensitive tensorial (e.g. octupolar) observables that are symmetric with respect to rotations equivalent to ellipticity reversals. This could play a role in the investigation of the combined influence of spin- and orbital-angular momentum (OAM) of light in chiral discrimination, which has recently emerged as an important topic. OAM beams intrinsically possess a spatially varying phase, which could wash out the conventional chiroptical response in multiple beam configurations, but leave NoDES signals unaffected. Similarly, NoDES could be used to in combination with quantum light. Bright squeezed vacuum states (BSV) \cite{iskhakov2012} are being used in a growing number of experiments to induce perturbations of strong-field processes \cite{rasputnyi2024,lemieux2025,tzur2025}. A second-harmonic BSV could similarly be used in the orthogonal two-color field configuration to ionize chiral molecules. However BSV fields intrinsically possess a random phase ambiguity of $\pi$ \cite{kern2025a}, which would wash out dichroic enantiosensitive signals. By contrast, the NoDES signal would survive, enabling to probe the influence of the quantum nature of light in chiral photoionziation.

We thank Franck Blais, Rodrigue Bouillaud, Romain Delos, Nikita Fedorov and Laurent Merzeau for technical assistance. We acknowledge Samuel Beaulieu, Baptiste Fabre and Nirit Dudovich for fruitful discussions. D.A. acknowledges funding from the Royal Society URF$\backslash$R$\backslash$251036. C. S. thanks Heriot-Watt University for funding through an EPSRC Doctoral Prize Fellowship. A. O. acknowledges funding from the Deutsche Forschungsgemeinschaft (DFG, German Research Foundation) 543760364. This work is part of the ULTRAFAST and TORNADO projects of PEPR LUMA and was supported by the French National Research Agency, under grant under ANR-21-CE30-038-01 (Shotime), and as a part of the France 2030 program, under grants ANR-23-EXLU-0002 and ANR-23-EXLU-0004. We acknowledge the financial support of the IdEx University of Bordeaux / Grand Research Program ``GPR LIGHT''. This project has received funding from European Union's Horizon 2020 research and innovation programme under grant  agreement no. 871124 Laserlab-Europe. The ELI ALPS project (GINOP-2.3.6-15-2015-00001) is supported by the European Union and co-financed by the European Regional Development Fund. 

\nocite{apsrev4-2Control}

\bibliographystyle{apsrev4-2}
\bibliography{biblio_full.bib}

\section{End Matter}

\begin{figure*}
	\centering
	\includegraphics[width=0.8\linewidth]{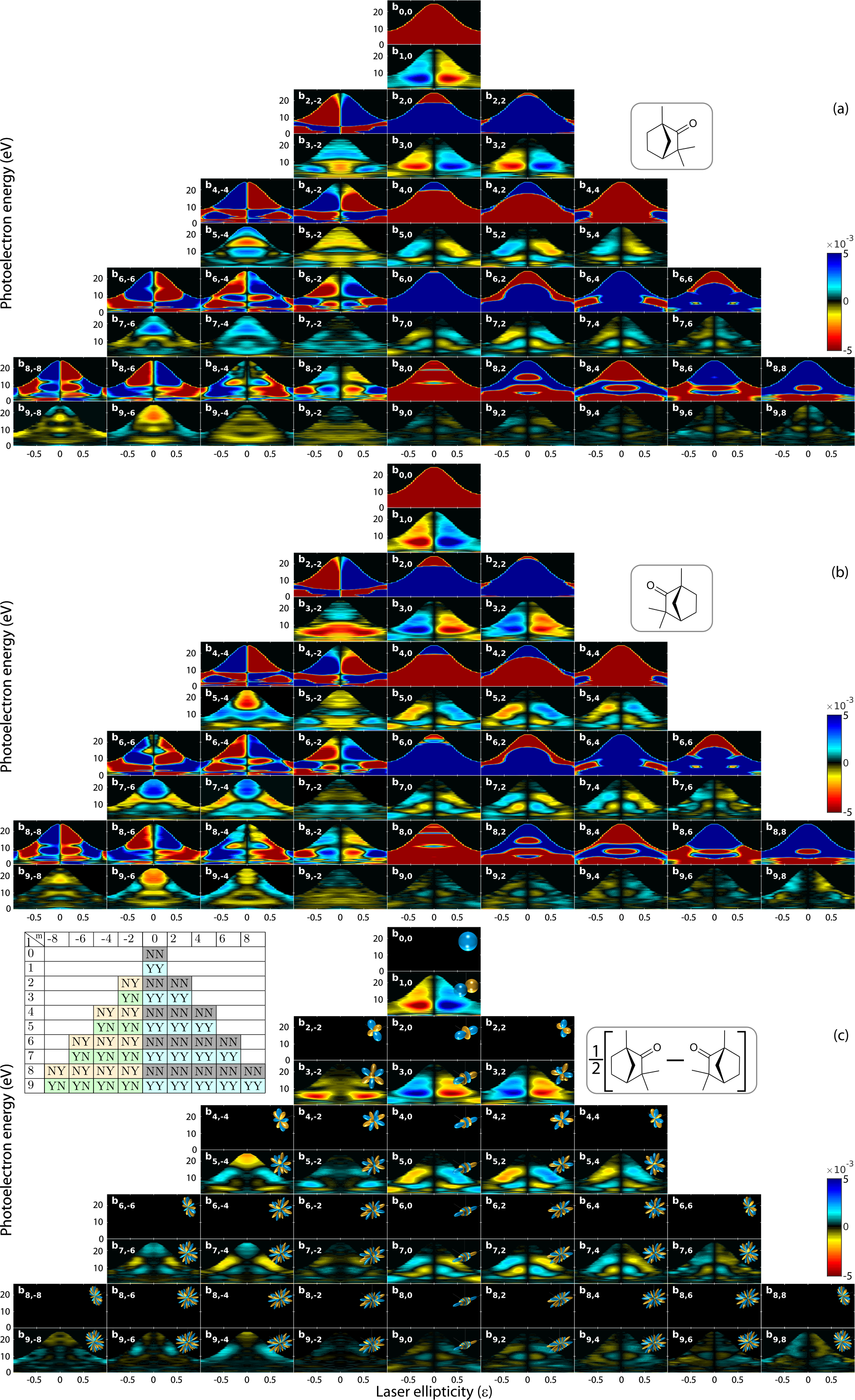} 
	\caption{Spherical harmonic decomposition [Eq. (\ref{eq:PAD})] of the photoelectron angular distributions obtained by photoionizing (+)-fenchone (a) and ($-$)-fenchone (b) with a 1030 nm laser field at  $2.5\times10^{13}$ $\mathrm{W/cm^{2}}$, as a function of the laser ellipticity. The decomposition of the enantiodifferential, forward-backward antisymmetric signal is shown in (c). The inset table recaps the enantiosensitivity (Yes=Y/No=N) and dichroism (Y/N) of the different coefficients.}
	\label{FigSH}
\end{figure*}

\begin{figure*}
	\centering
	\includegraphics[width=0.8\linewidth]{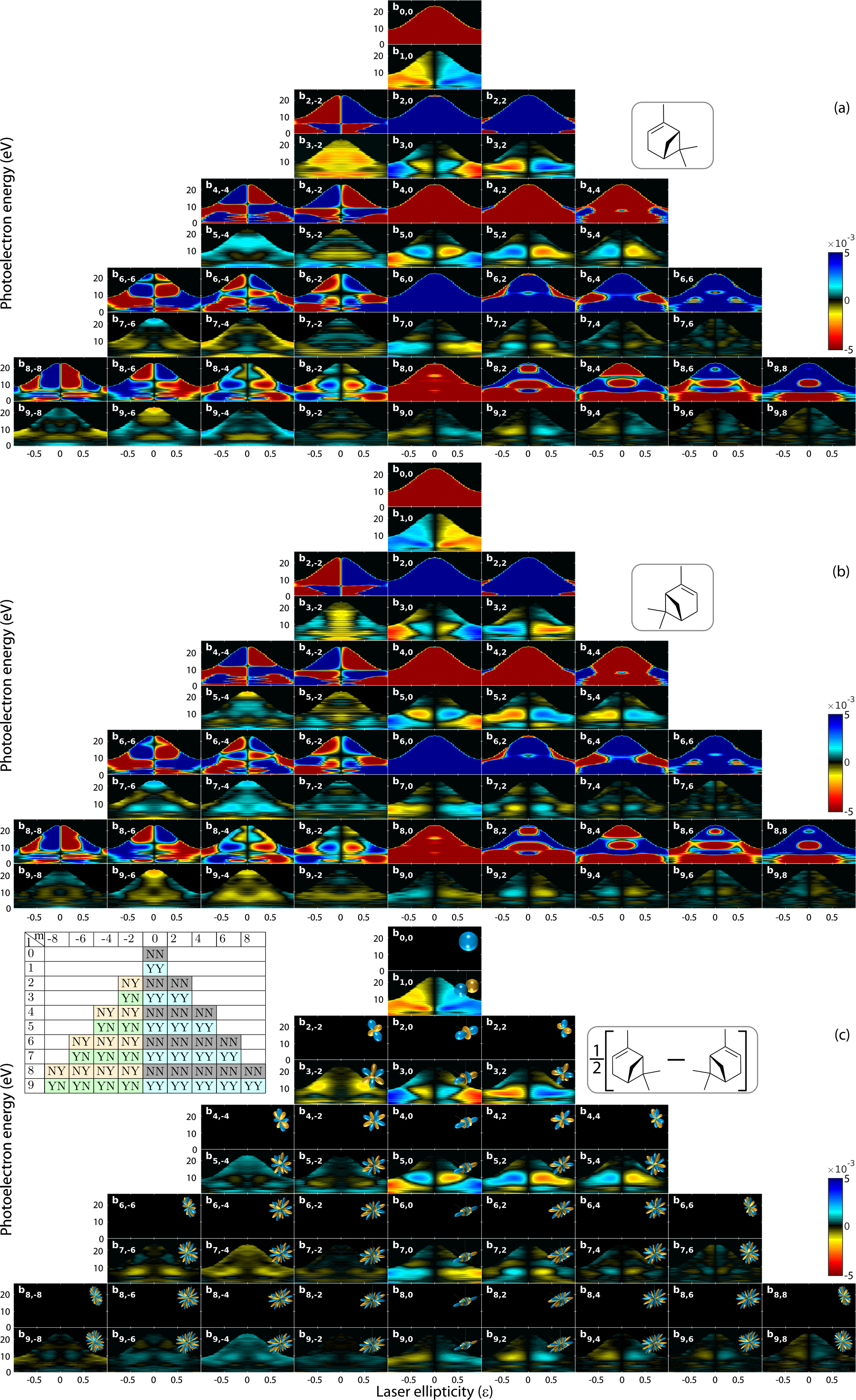} 
	\caption{Spherical harmonic decomposition [Eq. (\ref{eq:PAD})] of the photoelectron angular distributions obtained by photoionizing (+)-$\alpha$-pinene (a) and ($-$)-$\alpha$-pinene (b) with a 1030 nm laser field at  $2.5\times10^{13}$ $\mathrm{W/cm^{2}}$, as a function of the laser ellipticity. The decomposition of the enantiodifferential, forward-backward antisymmetric signal is shown in (c).  The inset table recaps the enantiosensitivity (Yes=Y/No=N) and dichroism (Y/N) of the different coefficients.}
	\label{FigSHPinene}
\end{figure*}

\end{document}